\RequirePackage{fix-cm}
\documentclass[twocolumn]{svjour3}          
\smartqed  
\usepackage{graphicx}
\usepackage{amssymb}
\usepackage{amsmath}
\usepackage{cases}
\usepackage{mathrsfs, bbm}
\usepackage{amsfonts}
\usepackage{epsfig}
\usepackage{booktabs}
\usepackage{upgreek}
\usepackage{graphicx,tabularx,multirow, color}
\usepackage[active]{srcltx}
\usepackage{pifont}

\makeatletter
\newcommand \rank {\mathop {\operator@font rank} }
\newcommand \level {\mathop {\operator@font level} }
\newcommand \sons {\mathop {\operator@font sons} }
\newcommand \dist {\mathop {\operator@font dist} }
\newcommand \diam {\mathop {\operator@font diam} }
\newcommand {\diag} {\mathop {\operator@font diag} }

\newcommand{\R}{\mathbb{R}}

\newcommand{\diff}{\,\mathrm{d}}
\newcommand{\bx}{\mathbf{x}}
\newcommand{\by}{\mathbf{y}}

\def \Os{\mathcal {O}}

\newcommand{\bbu}{\mathbf{u}}

\newcommand{\bbt}{\mathbf{t}}
\newcommand{\bbA}{\mathbf{A}}

\newcommand{\bbb}{\mathbf{b}}
\newcommand{\bba}{\mathbf{a}}
\newcommand{\bbT}{\mathbf{T}}
\newcommand{\bbU}{\mathbf{U}}

\def \ew{\mathrm {ew}}

\makeatother

\begin{document}

\title{Efficiency improvement of the frequency-domain BEM for rapid transient elastodynamic analysis 
}


\author{Jinyou Xiao         \and
        Wenjing Ye \and
        Lihua Wen 
}


\institute{J. Xiao \at
              College of Astronautics, Northwestern Polytechnical University,
Xi'an 710072, China \\
              \email{xiaojy@nwpu.edu.cn}           
           \and
           W. Ye \at
              Department of Mechanical Engineering, The Hong Kong University of Science and Technology, Clear Water Bay, Kowloon, Hong Kong\\
              \email{mewye@ust.hk}
}

\date{Received: date / Accepted: date}

\maketitle

\begin{abstract}

The frequency-domain fast boundary element method (BEM) combined with the exponential window technique leads to an efficient yet simple method for elastodynamic analysis. In this paper, the efficiency of this method is further enhanced by three strategies. Firstly, we propose to use exponential window with large damping parameter to improve the conditioning of the BEM matrices. Secondly, the frequency domain windowing technique is introduced to alleviate the severe Gibbs oscillations in time-domain responses caused by large damping parameters. Thirdly, a solution extrapolation scheme is applied to obtain better initial guesses for solving the sequential linear systems in the frequency domain.
Numerical results of three typical examples with the problem size up to 0.7 million unknowns clearly show that the first and third strategies can significantly reduce the computational time. The second strategy can effectively eliminate the Gibbs oscillations and result in accurate time-domain responses.

\end{abstract}

\section{Introduction}
\label{intro}

Transient analysis is encountered in almost every field in science and engineering. For most real-world problems, numerical simulation is the only viable approach for obtaining accurate solutions. The boundary element method (BEM) has been well-recognized as a powerful numerical method to treat wave
propagation problems.
Its reduced dimensionality, high accuracy and its ability in capturing rapid transitions and
steep gradients of the fields have made this method particularly suitable for problems with complex
geometry, for example, dynamic analysis of porous
solids and fracture analysis.
In view of the increasingly complex systems that we have encountered
nowadays, the development of efficient BEM for large-scale wave propagation simulations has become indispensable.

According to the different approximation solution strategies in time space, the BEM for treating elastodynamic wave propagation problems generally follows two approaches, namely, time-domain approaches and frequency-domain approaches (see e.g. the reviews by Beskos \cite{Bde97} and
Costabel \cite{CM04}).
Time-domain approaches can be further classified into time-stepping methods and the space-time integral equation method. In these methods the physical problems are directly solved in the real time domain, thus one can observe the phenomenon as it evolves.
However, such methods require an adequate choice of the time step size. An improper
time step could lead to instability or numerical damping. For recent development of time-domain methods, see e.g. \cite{SA97,BS12,ADF12}.

In the frequency-domain approach, the frequency-domain boundary integral equation is established and solved at a series of sampling frequencies. The results are then transformed back into time domain by inverse Laplace transform which can be numerically evaluated by the discretized Fourier transform (DFT) \cite{AM87,PTB05,PGS10}. This approach is attractive for transient analysis because it is: (1) stable in obtaining time-domain solutions, (2) simple in implementation and (3) suitable for parallel solution. In addition, one notes that the frequency-domain problems by itself already has a broad range of applications such as harmonic wave propagation in solids \cite{Bde97}, air damping on micro resonators \cite{DY04}, heat transfer \cite{SST12}, to name a few.

There are two issues need to be dealt with when computing the transient response using the inverse Fourier transform. The first issue is regarding to the discretization and truncation errors introduced by the DFT \cite{MR08}. The discretization in DFT causes wrap-aroud (or aliasing) error. For damped systems this error can be removed by adding trailing zeroes to damp out the free vibration before the end of the period. However, this scheme is inefficient for low damped systems, and more importantly, it does not work for non-dissipative systems such as the elastic solids considered in this work. To circumvent this problem, the exponential window method (EWM) \cite{KR92,Humar02} was employed in the BEM transient elastodynamic analysis \cite{PGSG11,XYCZ12}. This method adds an artificial damping to the system to effectively damp out the free vibrations. By properly choosing the damping parameter, denoted by $\eta$, in the EWM, the response period $T$ can be arbitrarily set independent of the actual time-domain response. In addition, it has been found that a large $\eta$ can improve the conditioning of the BEM coefficient matrix. However one cannot arbitrarily increase $\eta$  because it also amplifies the truncation error of DFT and therefore causes unacceptable Gibbs oscillations in the later time response. In this paper, a frequency-domain windowing technique is introduced to eliminate the Gibbs oscillations, and consequently accurate time response can be obtained in all simulation periods. The frequency-domain windowing technique in combine with the EWM leads to a simple but efficient modified Fourier transform approach for BEM elastodynamic transient analysis.

The second issue concerns the computational efficiency. In the frequency-domain approach the transient problem is transformed into $N_{\omega}$ independent time-harmonic problems. The computational cost grows almost linearly with $N_{\omega}$. For many practical problems  $N_{\omega}>100$ sampling frequencies are
often required to obtain accurate time-domain solutions.
This implies a huge computational cost because solving one frequency-domain eqation in a three-dimension domain using classic BEM is already computationally very costly.
To resolve this issue, a series of work has been recently published on the development of efficient integral equation approaches for frequency-domain elastodynamics. Examples include, but are not limited to, fast multipole \cite{SBD08,TC09,CBS08}, $\mathcal {H}$-matrix \cite{BA10} and precorrected-FFT (pFFT) accelerated BEM \cite{YZY10}, as well as efficient BEM with adaptive cross approximation \cite{MS08}. In this paper, the pFFT technique is employed to accelerate the frequency-domain BEM analysis \cite{XYCZ12} due to its ease in implementation and high computational efficiency.

A key factor that affects the computational cost of the frequency domain approach is the number of iterations required to solve each linear system. In order to achieve further speedup, a least square extrapolation method is proposed to obtain better initial guesses for the iterative solution procedure.

The basic setting of the frequency-domain approach for BEM transient analysis is briefly recalled in Section \ref{S-2}. The modified Fourier transform method for transient elastodynamic anaylsis is proposed in Section \ref{S-MFT}. In Section \ref{S-RSLS} the precorrected-FFT and the extrapolation method for initial guess are presented. The performance of the methods put forward in this paper is tested by three examples in Section \ref{S-NE}, and a summary is drawn in Section \ref{S-Conclution}.

\section{Frequency domain BEM for transient analysis}\label{S-2}

\subsection{Frequency domain boundary element method}
The first step in the frequency-domain approach for transient elastodynamic analysis is to transform the time domain  into the frequency domain via Fourier transform. Let $\Omega \in \R^3$ denote the region of space occupied by a three-dimensional
elastic solid with isotropic constitutive properties
defined by Lam\'e constants $\lambda$ and $\mu$, Poisson's ratio $\nu$ and mass density $\rho$. In a mixed boundary value problem, the boundary $\Gamma = \partial \Omega$ consists of a displacement boundary $\Gamma_u$ and a traction boundary $\Gamma_\sigma$ such that $\Gamma = \Gamma_u \cup \Gamma_\sigma$. The displacement governing equation for linear elasticity in the frequency domain is formulated as
\begin{equation}\label{governing-pde}
\begin{aligned}
\mu \Delta u(\bx, \omega) + (\lambda+\mu) \nabla \nabla \cdot u(\bx, \omega) &= \rho \omega^2 u(\bx, \omega), &\,& \bx \in \Omega, \\
u(\bx, \omega) &= \bar{u}(\bx, \omega), &\,& \bx \in \Gamma_u,\\
\sigma(\bx, \omega) \equiv (\mathscr{T}_{\bx}u)(\bx, \omega) &= \bar{\sigma}(\bx, \omega), &\,& \bx \in \Gamma_\sigma,\\
\end{aligned}
\end{equation}
where, $\omega$ is the circular frequency; $u=(u_1, u_2, u_3)$ is the displacement in the frequency domain; $\nabla$ and $\Delta$ denote the Nabla and the Laplace operators, respectively; $\mathscr{T}_{\bx}$ is the traction operator defining
the stress-strain relation based on Hooke's law; $\bar{u}$ and $\bar{\sigma}$ are prescribed boundary displacement and traction on $\Gamma_u$ and $\Gamma_\sigma$, respectively.

The boundary integral equation corresponding to equation \eqref{governing-pde} is formulated as
\begin{equation}\label{bie}
\begin{aligned}
& c_{ij}(\bx) u_j(\bx, \omega) + (\mathrm{P.V.}) \int_{\Gamma} T_{ij} (\bx,\by; \omega) u_j(\by, \omega) \diff \Gamma_{\by}\\
&= \int_{\Gamma} U_{ij} (\bx,\by; \omega) \sigma_j(\by, \omega) \diff \Gamma_{\by}, \quad \bx \in \Gamma
\end{aligned}
\end{equation}
where, (P.V.) indicates a Cauchy principal value (CPV) of the singular integral; the free-term $c_{ij}(\bx)$ is equal to $0.5\delta_{ij}$ for a smooth boundary at $\bx$; $U_{ij} (\bx,\by; \omega)$ and $T_{ij} (\bx,\by; \omega)$ denote the $j$th components of the displacement and
traction generated at $\by$ by a unit point force applied
at $\bx$ along the direction $i$, given by
\begin{equation}\label{Uij}
U_{ij} (\bx,\by; \omega) = {1 \over 4\pi \mu} \left[\varphi(r) \delta_{ij} - \psi r_i r_j \right],
\end{equation}
\begin{equation}\label{Tij}
T_{ij} (\bx,\by; \omega) = G_{ikj}(\bx,\by; \omega) n_k(\by),
\end{equation}
where,
\begin{equation}\label{Gikj}
\begin{aligned}
G_{ikj}(\bx,\by; \omega) = {1 \over 4\pi r} & \bigg[ \Big( {\diff \varphi \over \diff r} - {\psi \over r}  \Big) \Big(\delta_{ij} r_k + \delta_{ik} r_j \Big)\\
& + {2\over r^2} \Big(  {2 \psi \over r} - {\diff \psi \over \diff r} \Big)  r_i r_j r_k \\
& + \bigg( {c_p^2 \over c_s^2} \Big( {\diff \varphi \over \diff r} - {\diff \psi \over \diff r} - {2\psi \over r} \Big)\\
& - 2 \Big( {\diff \varphi \over \diff r} - {\diff \psi \over \diff r} - {\psi \over r} \Big)   \bigg) \delta_{jk} r_i \bigg];
\end{aligned}
\end{equation}
\begin{equation*}
\begin{aligned}
\varphi(r) &= {1 \over r} \Big[ \Big( 1 + {\iota \over k_s r}  - {1 \over k_s^2 r^2} \Big) e^{\iota k_s r}\\
  &\qquad\quad - \Big( {\iota \over k_p r}  - {1 \over k_p^2 r^2} \Big) e^{\iota k_p r} \Big],\\
\psi(r) &= {1 \over r} \Big[ \Big( 1 + {3\iota \over k_s r}  - {3 \over k_s^2 r^2} \Big) e^{\iota k_s r}  \\
&\qquad\quad - \Big( 1+ {3\iota \over k_p r}  - {3 \over k_p^2 r^2} \Big) e^{\iota k_p r} \Big];
\end{aligned}
\end{equation*}
$n(\by) = (n_1, n_2, n_3)$ is the unit normal to $\Gamma$ at $\by$ directed outwards of $\Omega$;
$$
r = ||\bx - \by||, \quad r_i = x_i - y_i;
$$
$k_s$ and $k_p$ are the respective wavenumbers of S and P elastic
waves, so that
$$
k_s = {\omega \over c_s}, \quad  k_p = {\omega \over c_p}; \quad c_s
= \sqrt{\mu \over \rho}, \quad c_p = \sqrt{\lambda + 2\mu \over
\rho}.
$$

The boundary integral equation \eqref{bie} is numerically solved using a surface discretization with $N_{\rm{e}}$ triangular boundary elements. Piecewise-constant approximation is employed in this work.
A standard collocation approach yields a discretized system of the BIE
\eqref{bie} of the form
\begin{equation}\label{Tu_Ut}
\bbT \bbu = \bbU \bbt,
\end{equation}
where matrices $\bbT$ and $\bbU$ with dimensions of $N$ by $N$, $N=3N_{\rm{e}}$, consist of boundary integrals of kernel functions $T_{ij}$ and $U_{ij}$ on each element respectively; the $N$-vectors $\bbu$ and $\bbt$ contain the nodal displacement and traction at each element. Note that the contribution of the free-term $c_{ij}(\bx)$ has been absorbed into the matrix $\bbT$.
By enforcing the boundary conditions, the linear system \eqref{Tu_Ut} becomes
\begin{equation}\label{Aa_b}
\bbA(\omega) \bba(\omega) = \bbb(\omega),
\end{equation}
where, the $N$-vector $\bba$ collects the unknown nodal displacement and traction components and can be obtained by solving the above linear system. All the matrix and vectors in \eqref{Aa_b} are functions of frequency $\omega$. The computational cost for an iterative solver is $\Os(N^2)$. This can be greatly reduced by using the pFFT method in section \ref{S-RSLS-pFFT}.

\subsection{Conventional frequency-domain approach} \label{S-2-FDA}

Equation \eqref{governing-pde} or \eqref{bie} is solved at a series of discrete frequencies. The resulting displacements and tractions are then transformed back to the time domain by using the inverse DFT to obtain the time-domain response.

Let $\Delta \omega$ and $\Delta t$ be the circular frequency and time resolutions, respectively; and $T$ be the time period of the transient response. Given the number of sampling points $N_{\rm{\omega}}$, one has the basic relations
$T = {2\pi / \Delta \omega}$ and $\Delta t = {T / N_{\rm{\omega}}}$. Note that the Nyquist frequency $f_{\rm{Nyq}} = N_{\rm{\omega}} \Delta \omega /(4 \pi)$ should be chosen such that responses corresponding to frequencies higher than $f_{\rm{Nyq}}$ are
insignificant and can thus be discarded.

The procedure for obtaining the transient responses using the frequency-domain approach can be summarized as:
\begin{enumerate}
  \item Determine a frequency resolution $\Delta \omega$ and the number of sampling points $N_{\rm{\omega}}$. $\Delta \omega$ needs to be small enough to minimize the loss of useful frequency information.
  \item Sample boundary condition $P(t)$, which could be the given boundary displacement or the traction, and perform DFT to obtain the frequency-domain boundary conditions $\hat P(\omega)$ at $N_{\rm{\omega}}$ frequencies $\{\omega_k= k \Delta \omega, \, (k=0, 1, \cdots, N_{\rm{\omega}}-1)\}$,

  \begin{equation}\label{DFT}
  \hat P(\omega_k) = {1 \over N_{\omega}} \sum^{N_{\omega} - 1}_{n=0} P(n \Delta t) e^{ -2\pi \iota n k /N_{\omega} }.
  \end{equation}

  \item Conduct analysis for the first $({N_{\rm{\omega}} \over 2}+1)$ frequencies, that is, $\{\omega_k = k \Delta \omega, \, (k=0, 1,  \cdots,  N_{\rm{\omega}}/2)\}$, to obtain the frequency-domain response $\hat R(\omega)$ for the first $({N_{\rm{\omega}} \over 2}+1)$ frequencies.  The responses $\hat R(\omega)$ corresponding to the last $({N_{\rm{\omega}} \over 2}-1)$ can be determined utilizing the conjugate symmetric property of the DFT as,
      $$
      \hat R(k) = \hat R^*(N_{\rm{\omega}}-k), \quad k={N_{\rm{\omega}}/ 2}+1, \cdots,  N_{\rm{\omega}}-1,
      $$
      where, $\hat R(k) = \hat R(k \Delta \omega)$ and $\hat R^*$ denotes the complex conjugate of $\hat R$.

  \item Perform IDFT for $\hat R(\omega)$ to obtain time-domain response $R(t)$, i.e., the desired boundary displacement and/or traction,
    \begin{equation}\label{IDFT}
  R(n \Delta t) = \sum^{N_{\omega} - 1}_{k=0} \hat R(\omega_k) e^{ 2\pi \iota n k /N_{\omega} }.
  \end{equation}

\end{enumerate}

Discretization and truncation errors arise when the forward and inverse Fourier transforms are numerically evaluated using DFT and IDFT in \eqref{DFT} and \eqref{IDFT}. The discretization error is also known as \emph{aliasing error}. To eliminate it, the time-domain signal is required to be zero for $t>T$. The truncation error is due to the truncation of the frequency spectrum. It causes high frequency oscillations near the discontinuities of the signals, known as \emph{Gibbs oscillations}.

In next sections, techniques for reducing the two errors are introduced. Consequently, a general and efficient modified Fourier transform (MFT) method for elastodynamic transient analysis is obtained. The MFT method itself is not new. It was first developed for analyzing power system transients by Mullineux's group about four decades ago; see \cite{MR08} and the references therein. In this paper, it is applied in elastodynamic transient analysis.

\section{Modified Fourier transform method}\label{S-MFT}

In the MFT method, two data processing techniques, one with respect to the time-domain data and another to the frequency-domain data, are introduced to reduce the discretization and truncation errors associated with the DFT and IDFT.

\subsection{Discretization error \& time-domain damping}\label{S-MFT-damping}
As aforementioned, the periodic nature of the DFT requires that the time-domain response tends to zero at the end of a period. This however often cannot be satisfied in the dynamic analysis using the frequency domain approach. A common way to circumvent this problem for a dissipated system is to add trailing zeroes to prolong the duration so that the free vibration can be damped out before the end of the period. Such an approach is clearly not applicable for a non-dissipated system. Another more general method suitable for all systems is the exponential window method (EWM) \cite{KR92,Humar02}. In this method, a complex frequency shift, amounting to an artificial damping, is introduced to efficiently damp out the free vibrations within the selected period. The actual response is obtained after a proper scaling of the damped response.

Recently, the EWM was used in 2D and 3D BEM elastodynamic transient analysis \cite{PGSG11,XYCZ12}. Accurate time domain response can be obtained by a proper choice of the damping parameter $\eta \geq 0$. To be specific, let $\hat u (\bx,t)$ be the displacement in the time domain and its Fourier transform is $u(\bx,\omega)$ in \eqref{governing-pde}. In applying the EWM, the displacement $\hat u$ is scaled by an exponential window function $e^{-\eta t}$ to obtain a damped displacement $\hat{u}_{\ew}$, i.e.,
\begin{equation}\label{scaled-u-TD}
\hat{u}_{\ew} (\bx, t) = e^{-\eta t} \hat u (\bx, t).
\end{equation}
Since $\hat{u}$ is a bounded function, the damped response $\hat{u}_{\ew}$ tends to be zero for large $t$. The Fourier transform of $\hat{u}_{\ew}$ is ${u}_{\ew} (\bx, \omega) =  u(\bx,\omega - \iota \eta)$,
meaning that ${u}_{\ew}$ can be obtained by simply replacing the frequency $\omega$ in $u$ by $\omega - \iota \eta$, where $\iota = \sqrt{-1}$. To obtain the transient response, one needs to solve the problem at a series of complex frequencies $\omega_k = k \Delta \omega - \iota \eta, k=0, 1,  \cdots,  N_{\rm{\omega}}/2$. Once the damped response $\hat{u}_{\ew}$ is obtained, the actual response $\hat u$ can be obtained as
\begin{equation}\label{u-remove-exp}
\hat u (\bx, t) = e^{\eta t} \hat{u}_{\ew} (\bx, t).
\end{equation}

The choice of the damping parameter $\eta$ determines the performance of the EWM. On one hand, it cannot be too large since the factor $e^{\eta t}$ in \eqref{u-remove-exp} acts as an amplifier which magnifies the cumulative errors due to the truncation of the frequency space which will be discussed in section \ref{S-MFT-truncation} and BEM error. On the other hand, given the factor $e^{-\eta t}$ is used to damp the time-domain response, it could be supposed that a high value for $\eta$ is required to sufficiently eliminate the aliasing error. A compromise between the two ends leading to the following rule-of-thumb for the selection of $\eta$ \cite{KR92}
\begin{equation}\label{ewm-k}
\eta = {\kappa \ln 10 \over T }= \kappa \Delta f \ln 10.
\end{equation}
This rule was used in the work described in \cite{PGSG11,XYCZ12}. It was found in \cite{XYCZ12} that in BEM analysis a choice of the coefficient $\kappa$ between $2 < \kappa < 3$ could often lead to good time-domain results.

In addition to reducing the aliasing error, another important effect of $\eta$ lies in that a positive $\eta$ can improve the conditioning of the system matrices in \eqref{Tu_Ut} and \eqref{Aa_b}. This is a nature result of the \emph{Gershgorin circle theorem}, because the exponential function $e^{-\eta t}$ reduces the absolute values of the non-diagonal entries of the matrices, thus making the eigenvalues more clustered. Hence, the damping factor $e^{-\eta t}$ actually acts as a perconditioner (called \emph{artifical damping perconditioner}, ADP, in this paper) whose performance can be enhanced by increasing $\eta$. As such, a larger $\kappa$ is always desirable.

Matrix preconditioning is a crucial problem in BEM analysis. This is particularly true for high frequency and multi-domain problems, in which the coefficient matrices tend to be much more ill-conditioned. Various preconditioners exist (e.g., \cite{INYN12,CSB12}) and can be used. However, it is well known that the performance of many existing preconditioners is generally problem-dependent and there always is a trade-off between their performance and computational cost. On the contrary, the ADP is simple yet efficient, because its implementation requires no extra computation and memory. Whereas, it can largely increase the convergence speed of the iterative solvers; see Section \ref{S-NE}.

Meanwhile, the ADP can be used in combination with other preconditioners. For instance, in this work a block diagonal preconditioner is used together with the time-domain damping technique described in this section to solve \eqref{Aa_b}.

\subsection{Truncation error \& frequency-domain windowing}\label{S-MFT-truncation}

It is observed that choosing $\kappa > 2$ in BEM analysis can cause unacceptable Gibbs oscillations in time-domain responses at later times \cite{XYCZ12}. Here, a frequency-domain data windowing technique is introduced to alleviate the the Gibbs oscillations. Consequently, a larger value of $\kappa$ can be set. This implies a considerable improvement in the conditioning of the BEM matrices when compared with $\kappa =2$; see evidences in Section \ref{S-NE}.

The implementation of this technique is rather simple \cite{MR08}. Before the damped frequency-domain response ${u}_{\ew} (\bx, \omega)$ is transformed back into the time domain using IDFT, it is first multiplied by a suitable data window $W(\omega)$; that is, the function $ W(\omega) {u}_{\ew} (\bx, \omega)$ is transformed in lieu of ${u}_{\ew} (\bx, \omega)$ itself. The function of the frequency-domain windowing is to reduce the Gibbs oscillations by averaging out the oscillations within one period.

There exist a number of window functions, see e.g., \cite{MR08,CC95,CC98}. As far as BEM elastodynamic analysis is concerned, we find that the Hanning and Blackman windows often yield satisfactory results. The discretized window functions are given by
\begin{equation*}
\begin{aligned}
  \text{Hanning:} & \; W(\omega_k) = 0.5 \left[ 1+ \cos \left( {2 \pi k \over N_{\omega}} \right) \right],\\
  \text{Blackman:} & \; W(\omega_k)= 0.42 + 0.5\cos \left( {2 \pi k \over N_{\omega}} \right) \\
  & \quad\qquad\qquad + 0.08 \cos \left( {4 \pi k \over N_{\omega}} \right),\\
  &\quad k=0,1,\cdots, N_{\omega}-1.
\end{aligned}
\end{equation*}

\subsection{A summary}

To summarize, the time-domain damping in conjunction with the frequency-domain windowing techniques leads to a MFT method for elastodynamic transient analysis. The basic implementation procedure of the MFT method is similar to that described in section \ref{S-2-FDA}. The distinction lies in that in the MFT method the sampling frequencies are $\omega_k = k \Delta \omega - \iota \eta$ instead of $\omega_k = k \Delta \omega$, and the DFT formula \eqref{DFT} and IDFT formula \eqref{IDFT} are replaced by
  \begin{equation}\label{DFT-MFT}
  \hat P(\omega_k) = {1 \over N_{\omega}} \sum^{N_{\omega} - 1}_{n=0} e^{-\eta n \Delta t} P(n \Delta t) e^{ -2\pi \iota n k /N_{\omega} }
  \end{equation}
and
  \begin{equation}\label{IDFT-MFT}
  R(n \Delta t) = e^{\eta n \Delta t} \sum^{N_{\omega} - 1}_{k=0}  W(\omega_k) \hat R(\omega_k) e^{ 2\pi \iota n k /N_{\omega} },
  \end{equation}
respectively.

\section{Rapid solution of linear systems}\label{S-RSLS}

The main computational cost for the frequency-domain approach is the solution of linear systems \eqref{Aa_b} for a sequence of sampling frequencies $\omega_k$, i.e., the parameterized linear systems
\begin{equation}\label{Aa_b-sequence}
\bbA_k \bba_k = \bbb_k, \quad k=0,1,\cdots, {N_{\rm{\omega}} \over 2},
\end{equation}
where, $\bbA_k = \bbA (\omega_k)$, $\bba_k$ and $\bbb_k$ are defined analogously. In this section, two methods are described to considerably reduce the computational cost, namely, a pFFT method for accelerating the BEM matrix-vector multiplication, and a \emph{frequency extrapolation} method for obtaining good initial guesses and thus reducing the total number of iterations in solving the sequential systems.

\subsection{PFFT accelerated elastodynamic BEM} \label{S-RSLS-pFFT}

The computational cost of a matrix-vector multiplication in traditional BEM is $\Os(N^2)$, which could be prohibitively expensive for large-scale elastodynamic simulations. Within the last two decades, significant progress has been made in the development of accelerated BEM  techniques. With the accelerated techniques, the computational complexity, including both the CPU time and memory usage, can be considerably reduced, therefore enabling BEM-based large-scale computation.

The first accelerated technique used in elastodynamic BEM is the FMM \cite{F98}, which is further developed by several groups; see, e.g., \cite{TC09,CBS08,TNK03}. In \cite{BA10} the $\mathcal {H}$-matrix method is applied in frequency-domain elastodynamic BEM. Recently the pFFT method was used to accelerate elastodynamic BEM analysis by the present authors \cite{XYCZ12}. The pFFT method is very easy to implement and to be parallelized, so a high computational efficiency can be achieved with ease. For more information about the pFFT technique and the details of its implementation, the readers are referred to \cite{XYCZ12}.

Theoretically, for problem domains with small surface-to-volume ratios such as porous solids, optimal order $\Os(N \log N)$ for CPU time and $\Os(N)$ for memory can be achieved by pFFT, which is comparable to the FMM and the $\mathcal {H}$-matrix method. For other problems, the CPU time of pFFT is between $\Os(N\log N)$ and $\Os(N^{1.5}\log N)$. In these cases, the FMM and the $\mathcal {H}$-matrix method would be advantageous.
However numerically, due to the small overhead, the actually computational cost of pFFT could be less than that of FMM and the $\mathcal {H}$-matrix method for problems with $N$ up to order $\Os(10^5)$; see \cite{XYCZ12} for comparisons.

\subsection{Solution extrapolation method} \label{S-RSLS-IG}

Iteratively solving each linear system in the sequence \eqref{Aa_b-sequence} needs an initial guess, denoted by $\bba_k^{(0)}$. Typically, $\bba_k^{(0)}$ is independently chosen as zero or one vectors. Here a least square extrapolation method is proposed to determine a better initial guess $\bba_k^{(0)}$ for the solution of $\bba_k$ based on the calculated solutions at previous frequencies in the sequence; this approach is termed as the solution extrapolation method (SEM).

The motivation is inspired by the fact that solution $\bba (\omega)$ depend on the frequency $\omega$. If $\bba (\omega)$ is a smooth function of $\omega$ and that the frequency step $\Delta \omega$ is sufficiently small, it should be possible to obtain $\bba_k^{(0)}$ by extrapolation from the previous $K$ solutions; that is,
\begin{equation}\label{a0}
\bba_k^{(0)} = \sum^K_{i=1} s_i \bba_{k-i} = \mathbf{X} \mathbf{s},
\end{equation}
where, $\mathbf{X}=( \bba_{k-1},\cdots, \bba_{k-K} )$, $\mathbf{s}=( s_{1},\cdots, s_{K} )^{\text{T}}$. Let $\mathbf{Y} = \bbA_k \mathbf{X}$.
Then the coefficient vector $\mathbf{s}$ can be obtained by minimizing the square error
\begin{equation}\label{lsem}
||  \bbA_k \bba_k^{(0)} - \bbb_k ||_2 = ||  \mathbf{Y} \mathbf{s} - \bbb_k ||_2.
\end{equation}
By computing the QR-decomposition $\mathbf{Y} = \mathbf{QR}$, one gets the least square solution of \eqref{lsem}
\begin{equation*}
\mathbf{s} = \mathbf{R}^{-1}\mathbf{Q}\bbb_k.
\end{equation*}
Thus, the initial guess is can be computed as
\begin{equation}\label{a0-solution}
\bba_k^{(0)} = \mathbf{X}\mathbf{R}^{-1}\mathbf{Q}\bbb_k.
\end{equation}
The main computational cost for evaluating $\bba_k^{(0)}$ is the computation of $K$ matrix-vector products $\bbA_k \mathbf{X}$.

Optimal value for the number of preceding solutions $K$ in \eqref{a0} clearly depends on the smoothness of the solutions $\bba (\omega)$ and thus is problem dependent. For the cases studied in this work, it is found that $K=3,4$ is optimal. Generally, a large $K$ does not improve the accuracy of the initial guess, because the solution $\bba (\omega)$ is often not smooth over a large frquency range and the frequency step $\Delta \omega$ may not be so small as required.

\section{Numerical examples}\label{S-NE}

In this section, three examples are selected to validate and demonstrate the performance of the MFT method in Section \ref{S-MFT} and the SEM for initial guess in Section \ref{S-RSLS-IG}. The first example is a classical benchmark for transient elastodynamic analysis: a prismatic rod subject to a sudden applied uniform pressure at one end. The analytic solution is available and is employed to evaluate the accuracy and the efficiency of the method. The second case is an aluminum plate subject to an impact loading. This example is chosen to test the method for problems with a time-dependent arbitrary loading. Finally to test the ability of the method for solving large-scale problems, a unit cube containing 64 spherical cavities subject to a sudden applied uniform pressure is simulated.

In the following simulations, constant elements are used. The pFFT method in \cite{XYCZ12} is used to accelerate the BEM analysis. All the linear systems are preconditioned using the block diagonal preconditioner and solved by the generalized minimal residual method (GMRES) \cite{SS86} with convergence tolerance $10^{-5}$. Calculations are performed on a workstation with a Xeon 3.0GHz CPU and 30GB RAM.

\subsection{Example 1: prismatic rod} \label{S-S-NE1}

\begin{figure}[hbt]
\centering
\epsfig{figure=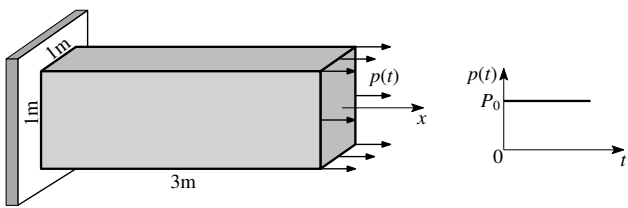,width=0.47\textwidth}
\caption{Prismatic rod subject to a step traction pulse}
\label{fig-rod}
\end{figure}

As the first test case, a prismatic rod with dimensions of $3\rm{m}\times 1\rm{m} \times 1\rm{m}$ as sketched in Figure \ref{fig-rod} is simulated. The rod is fixed at its left end and subject to a Heaviside traction $p(t)$ = $P_0 H(t)$ at its right end, where $P_0 = -10^6$ N/m$^2$ and $H(t)$ is the Heaviside function. By setting the Poisson's ratio of the material to zero, the 3-D problem is reduced to a 1-D problem, of which the analytical solution is available, see \cite{XYCZ12}. The Young's modulus $E$ and the density $\rho$ of
the material are set to be $E = 2.11\cdot 10^{11}$ Pa and
$\rho = 7.85 \cdot 10^{3}$ kg/m$^3$, respectively.

\subsubsection{Effects of the ADP and the SEM} \label{S-S-NE1-1}
The boundary of the prismatic rod is partitioned into 12800 triangular elements. The response period and the number of samplings are set to be $T = 0.0155$s and $N_{\rm{\omega}} = 256$, respectively. To test the effects of the damping coefficient $\kappa$ and the SEM on accelerating the convergence of the linear system solver, simulations are carried out under the three cases specified by the first three columns of Table \ref{tab-NE1-3cases}. For cases without using the SEM, the initial guesses are set to be zero vectors.

\begin{table}[hbt]
\begin{center}
\caption{Three test cases and their CPU times}\label{tab-NE1-3cases}
\begin{tabular*}{0.46\textwidth}{@{\extracolsep{\fill}}ccccc@{}}\toprule
Cases   & $\kappa$  & SEM        & Total its & Total time (h) \\
\hline
Case 1  & 2         &  \ding{55} & 29237 & 16.0    \\
Case 2  & 4         &  \ding{55} & 21758 & 14.4    \\
Case 3  & 4         &  \ding{51}, $K=4$ & 17562 & 12.1    \\
\bottomrule
\end{tabular*}
\end{center}
\end{table}

\begin{figure*}[hbt]
\centering
\epsfig{figure=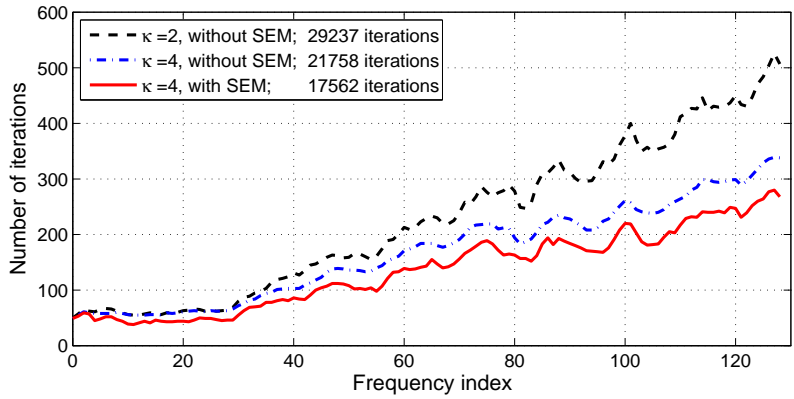,width=0.7\textwidth}
\caption{Total number of iterations in solving the linear systems to the sampling frequencies}
\label{fig-iterations}
\end{figure*}

The number of iterations for solving the sequential linear systems \eqref{Aa_b-sequence} corresponding to the first 129 sampling frequencies are plotted in Figure \ref{fig-iterations}. The total number of iterations and the total CPU time for solving all the 129 systems in each of the three cases are listed in the last two columns of Table \ref{tab-NE1-3cases}. One can see that by using a larger damping with $\kappa=4$, a 25\% reduction in the total number of iterations can be achieved when compared with the case of $\kappa=2$. By further using the SEM, the reduction is up to about 40\%. However, the reduction of the total CPU time is not as considerable as the number of iterations. This is because the matrix-vector multiplication in this problem is relatively fast and the preprocessing time (i.e., time for computing the near-field interations) occupies a large portion.

\subsubsection{Effects of the frequency windowing technique} As described in Section \ref{S-MFT-truncation}, the frequency windowing technique can be employed to eliminate the Gibbs oscillations in the time-domain response. Here this is verified using the results of case 3, i.e. $\kappa = 4$, in Table \ref{tab-NE1-3cases}. Our numerical experience shows that both the Blackman and Hanning windows obtains almost the same results. Thus only the results of Hanning window are presented here.

\begin{figure*}[hbt]
\centering \epsfig{figure=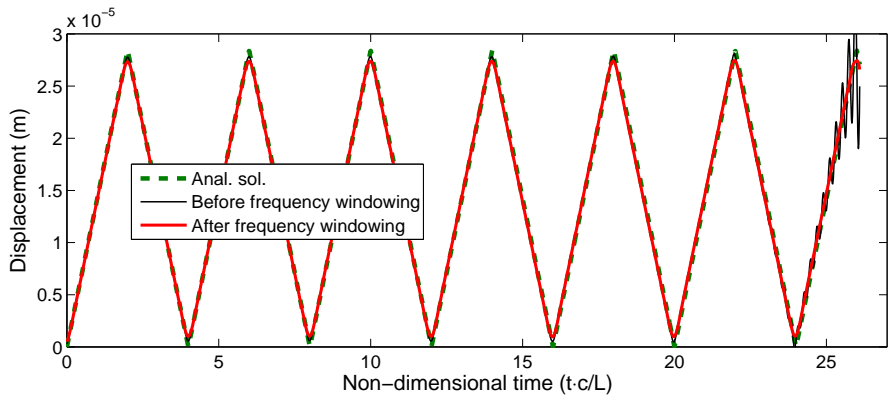,width=0.8\textwidth}
\caption{Displacement response at the free end of the rod} \label{fig-NE1-disp}
\end{figure*}

\begin{figure*}[hbt]
\centering \epsfig{figure=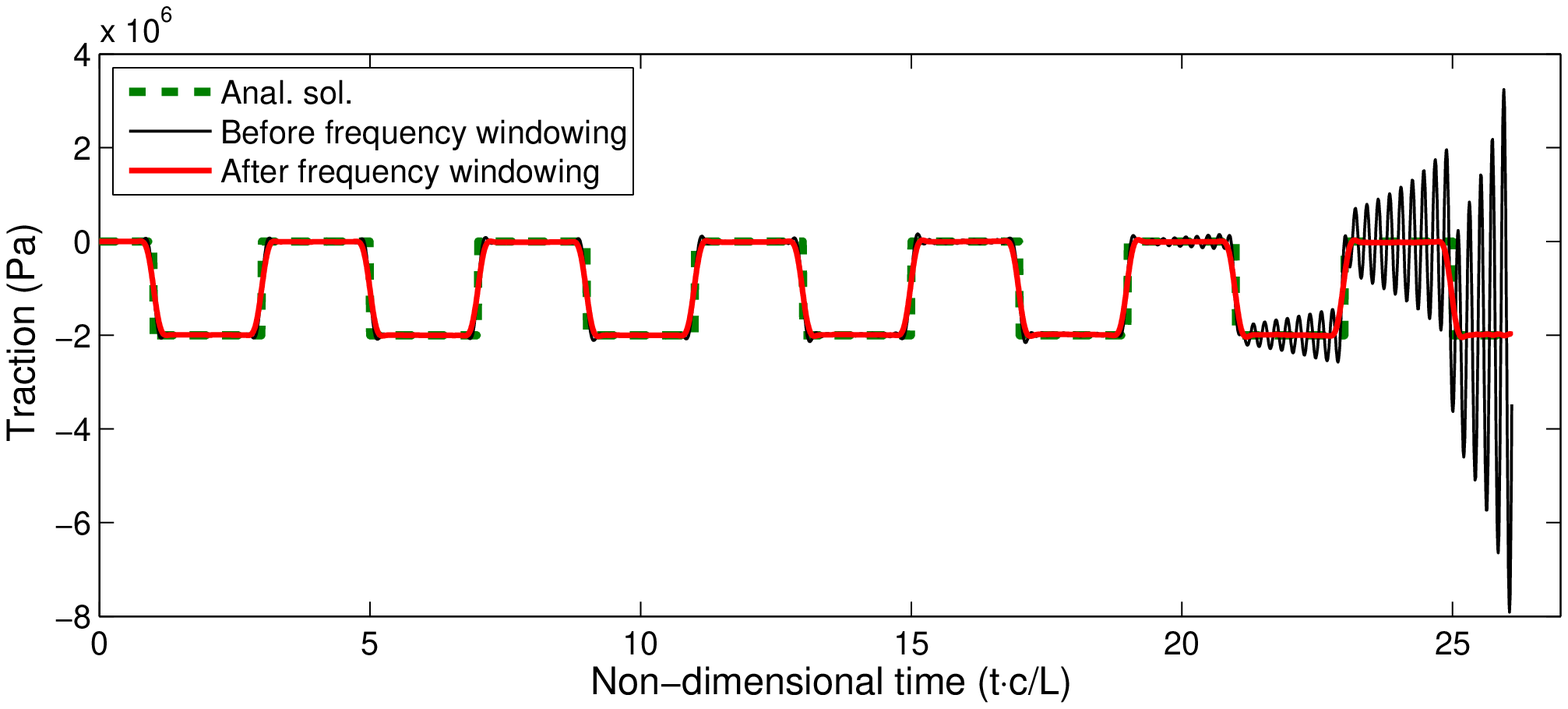,width=0.8\textwidth}\\
{\small (a) Total time response}\\
\centering \epsfig{figure=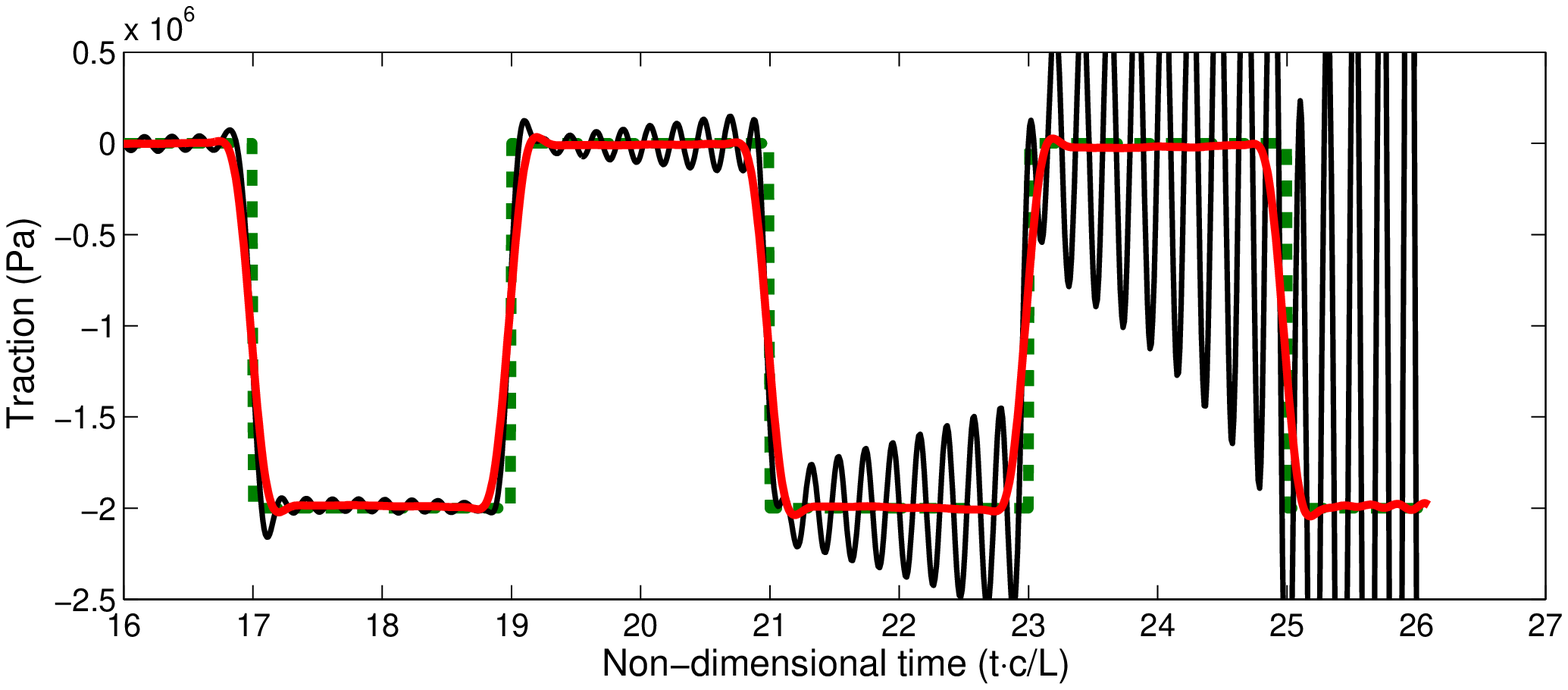,width=0.8\textwidth}\\
{\small (b) Local magnification}\\
\caption{Traction response at the fixed end of the rod} \label{fig-NE1-tract}
\end{figure*}

Figure 3 and 4 demonstrate both the analytical and simulated
displacements at the free end and the tractions at the fixed end as functions of time; the
non-dimensional time ($t\cdot c/L$, with $c$ and $L$ being the wave speed and length of the rod, respectively) is employed in the plots. It is observed that, before using the frequency windowing technique, both the displacement and traction responses show drastic oscillations in later times. The magnitude of the traction oscillations is about 4 times larger than the largest traction value; see Figure \ref{fig-NE1-tract} (a). But after the frequency windowing technique is used, the red curves obviously show that the oscillations are effectively removed and the resulting responses agree very well with the analytical solutions in all the simulation period; see Figure \ref{fig-NE1-tract} (b).

\subsubsection{Long time behavior of the method} It is well known that time-domain BEM analysis often suffers from either strong numerical damping or instabilities when the calculation time period is long. Here, the long-time behavior of the present frequency domain method is studied. The simulation conditions are the same as the case 3 in Section \ref{S-S-NE1-1}; but the response period and the number of samplings are set to be $T = 0.035$s and $N_{\rm{\omega}} = 512$, respectively. Thus, the corresponding time step is $\Delta t = 6.84\cdot 10^{-6}$s and 257 frequency domain BEM analyses are needed.

\begin{figure*}[hbt]
\centering \epsfig{figure=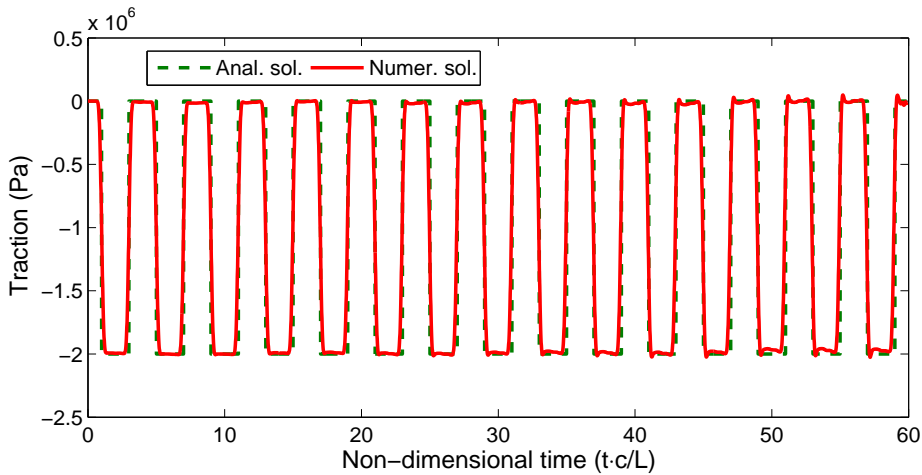,width=0.8\textwidth}\\
\caption{A long time traction response at the fixed end of the rod} \label{fig-NE1-tract-long}
\end{figure*}

The computed traction history is shown in Figure \ref{fig-NE1-tract-long}. Clearly, the proposed method shows a very nice behavior in the entire simulation period. Computed traction responses of similar problems can be found in Figure 12 of \cite{BMS12} and Figure 9 of \cite{ADF12}. It seems that the total numbers of time steps required in their work are much larger than that in the present simulation. For example, the number of time steps for obtaining Figure 9 of \cite{ADF12} is about $600/0.125 = 4800$, whereas in the present method only 257 frequency steps corresponding to 257 time steps are needed.

\subsection{Example 2: a plate with a hole subject to an impact loading} \label{S-S-NE2}

\begin{figure}[hbt]
\centering
\epsfig{figure=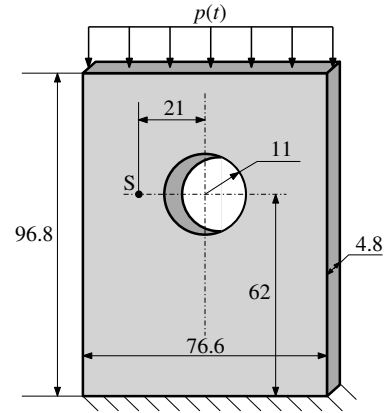,width=0.28\textwidth}
\caption{A plate with a hole (unit: mm)}
\label{fig-plate-hole}
\end{figure}

The performance of the present method is further demonstrated by a more realistic problem; i.e., an aluminum plate with a hole subject to a
vertical impact force as illustrated in Figure
\ref{fig-plate-hole}. The material
properties of the aluminum plate are: $E=69$ GPa, $\rho=2.7 \cdot
10^3$ kg/m$^3$ and $\nu=0.3$. For more details of the simulation conditions and the load history, the readers are referred to Section 4.2 of \cite{XYCZ12}. In the present computation, a mesh with 14072 triangular elements is used, $T=0.003$s, $N_\omega = 160$. The damping coefficient $\kappa$ is set to be 4, the SEM with $K=4$ is used. The Hanning window is employed for frequency windowing.

The responses of displacement and strain in the load-direction at points S (see Figure \ref{fig-plate-hole}) are computed using the present method and compared with the FEM and/or experiment results in Section 4.2 of \cite{XYCZ12}; see Figure \ref{fig-NE1-disp_strain}. By using the frequency windowing technique the Gibbs oscillations in both displacement and strain responses in the later times are effectively removed. The windowed results coincide very well with the FEM results. The agreement of the computed strain with the experimental measurement is also quite good overall.

The total CPU time for this simulation is about 5.4 hours. The memory usage is 840 MB, slightly more than the 723 MB in \cite{XYCZ12}, because here we use GMRES solver which is more memory consuming than the IDR(s) solver used in \cite{XYCZ12}.

\begin{figure*}[hbt]
\centering \epsfig{figure=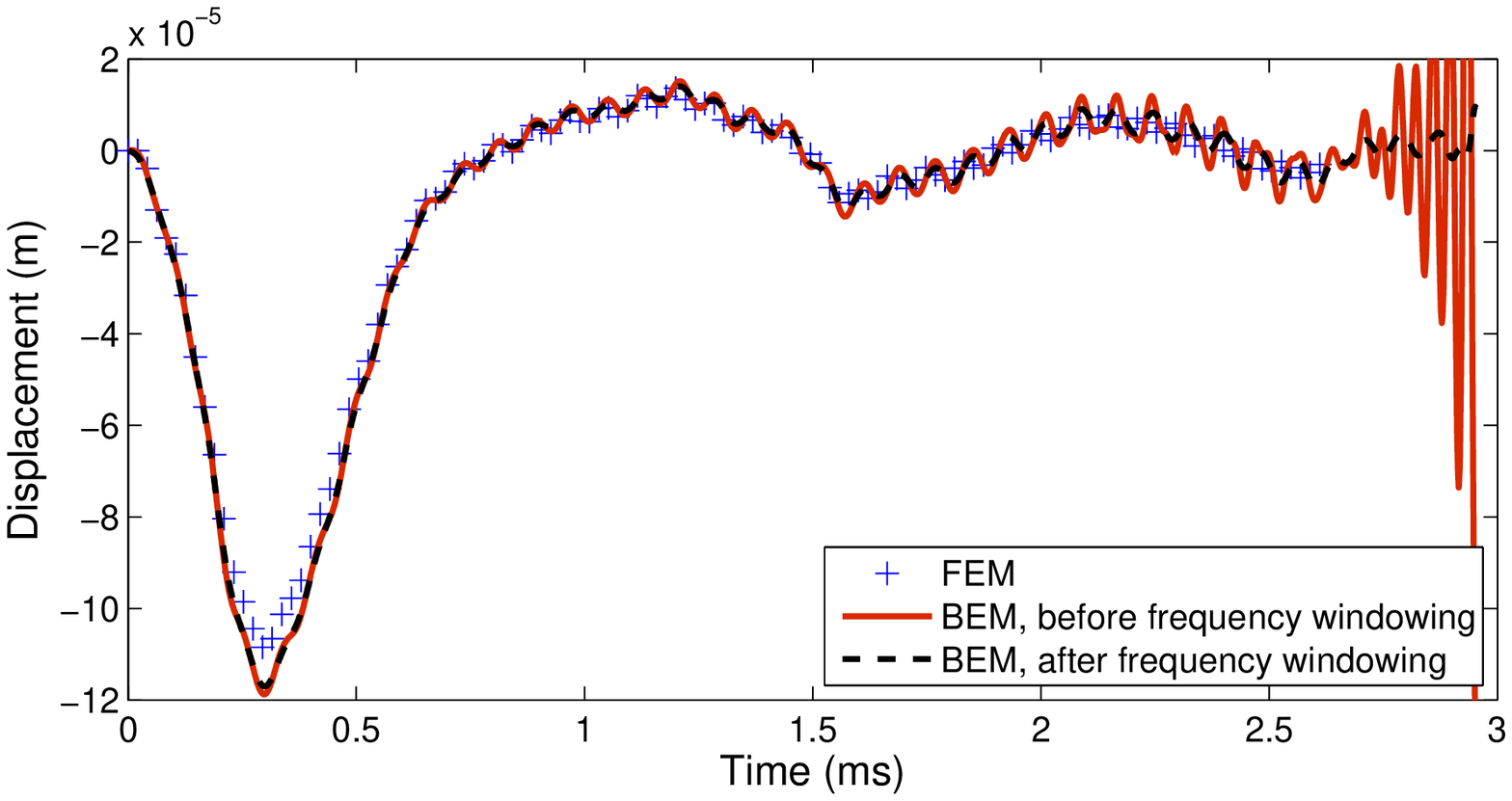,width=0.8\textwidth}\\
(a) Displacement
\end{figure*}
\begin{figure*}[hbt]
\centering \epsfig{figure=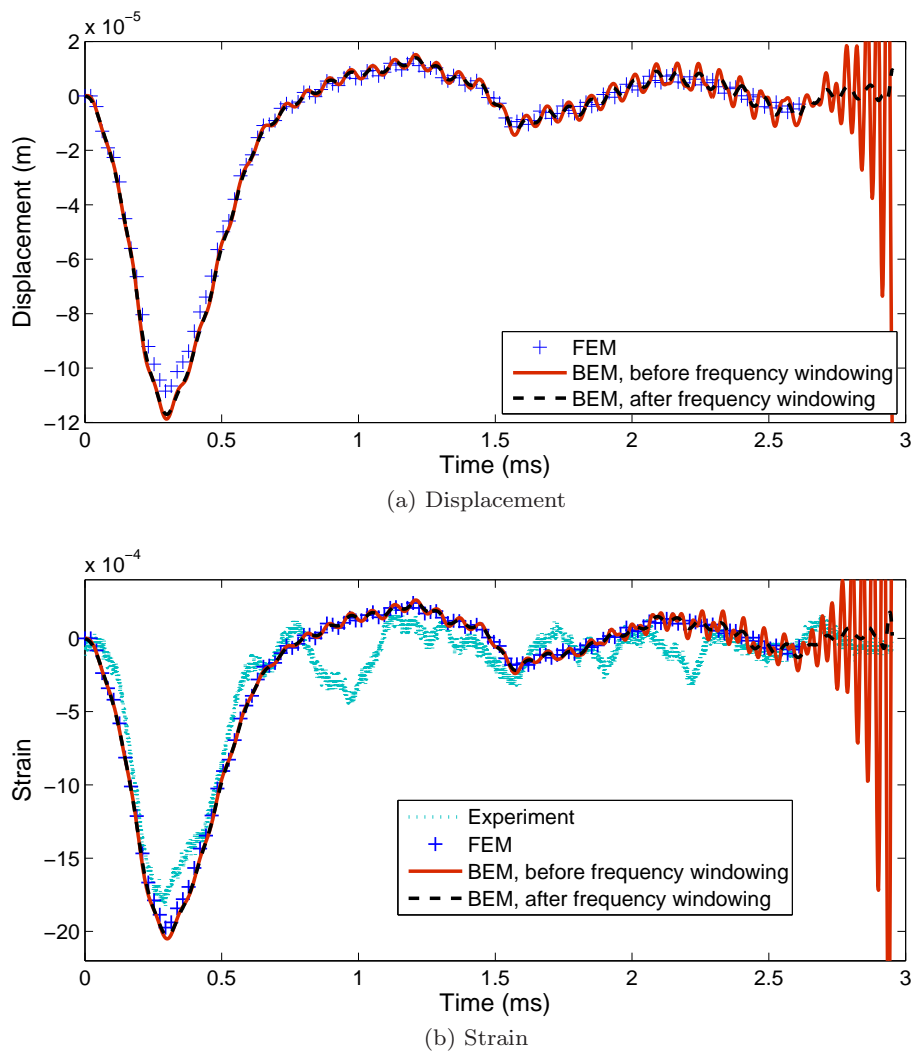,width=0.8\textwidth}\\
(b) Strain
\caption{The displacement and strain responses at point S} \label{fig-NE1-disp_strain}
\end{figure*}

\subsection{Example 3: a cube with 64 spherical cavities} \label{S-S-NE3}

\begin{figure}[hbt]
\centering
\epsfig{figure=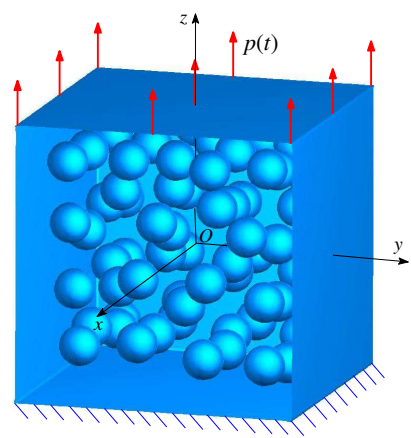,width=0.3\textwidth}
\caption{A unit cube containing 64 spherical cavities subjecting to a step load}
\label{fig-cube64cav}
\end{figure}

To demonstrate the performance of the presented method for solving large-scale elastodynamic wave propagation
problems, a model consisting of a solid cube with 64 identical spherical cavities
was built and simulated; see Figure \ref{fig-cube64cav}. The non-overlapped spherical cavities are
randomly distributed inside the cube and their radius is 0.072 m. The cube is fixed at the bottom
surface ($z = -0.5$) and is subject to a uniform step traction $p(t)$ at the upper surface ($z=0.5$). The load $p(t)$ is the same as that in Section \ref{S-S-NE1}. All
other surfaces are traction free. The material is aluminum and the properties are the same as those defined in
Section \ref{S-S-NE2}.
This example was also used in \cite{XYCZ12}, but here a finer mesh with $N_{\mathrm{e}}= 231422$ elements is used; hence, the total degrees of freedoms is nearly 0.7 million. The response period and the sampling size are set to be $T=0.008$s and $N_\omega = 160$, respectively.
Damping coefficient $\kappa = 4.0$ and the SEM with $K=4$ are used. The Hanning window is employed in frequency windowing.

The simulation takes about 101 hours, and consumes 21 GB memory. The CPU time still
seems to be quite large. Further reduction of computational expense can be achieved by employing a more effective pre-conditioning method and by parallel computing. The time responses of the displacement and traction in $z$-direction at four
boundary points are plotted in Figure \ref{fig-NE3-disp_tract}. Compared with the 1D solutions of the solid rod presented
in Section \ref{S-S-NE1}, the oscillation patterns of the displacement and traction seem reasonable.

\begin{figure*}[hbt]
\centering \epsfig{figure=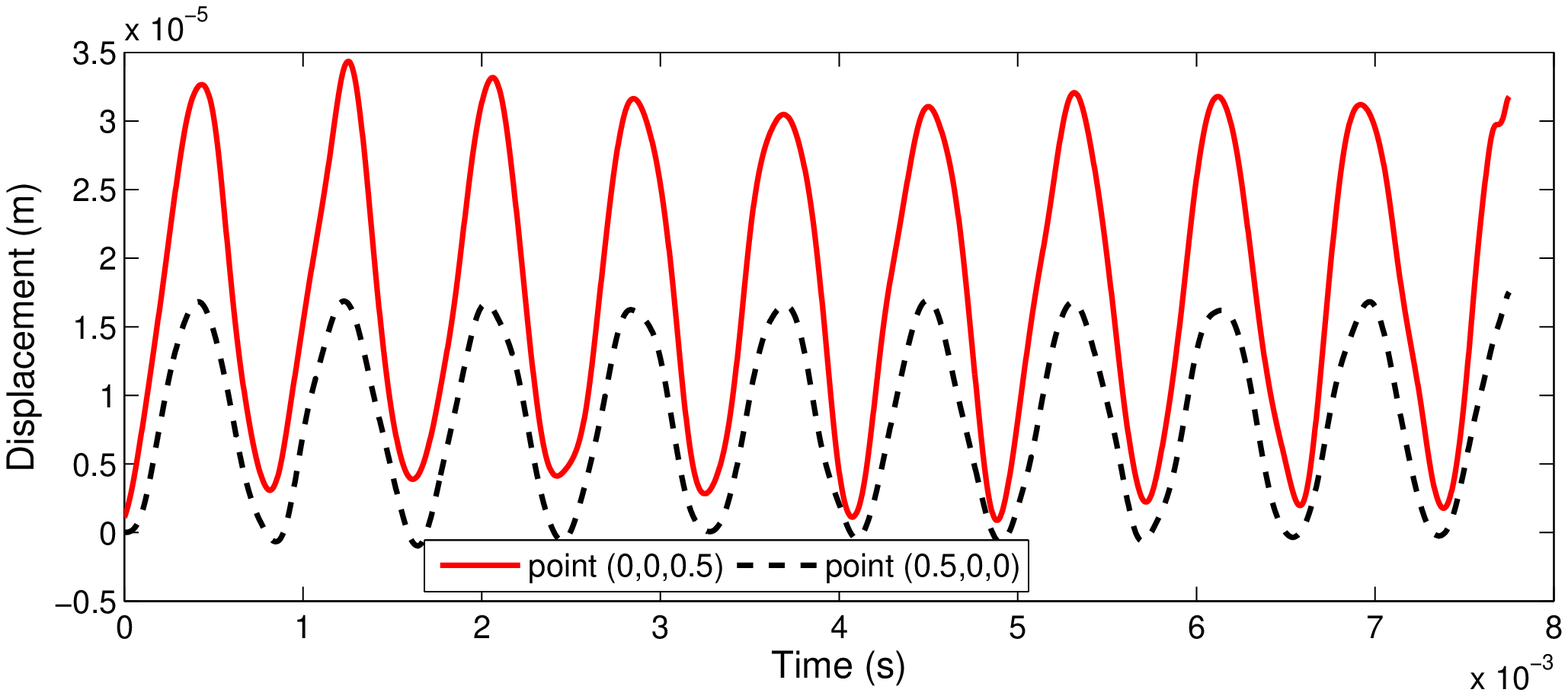,width=0.8\textwidth}\\
(a) Displacement
\end{figure*}
\begin{figure*}[hbt]
\centering \epsfig{figure=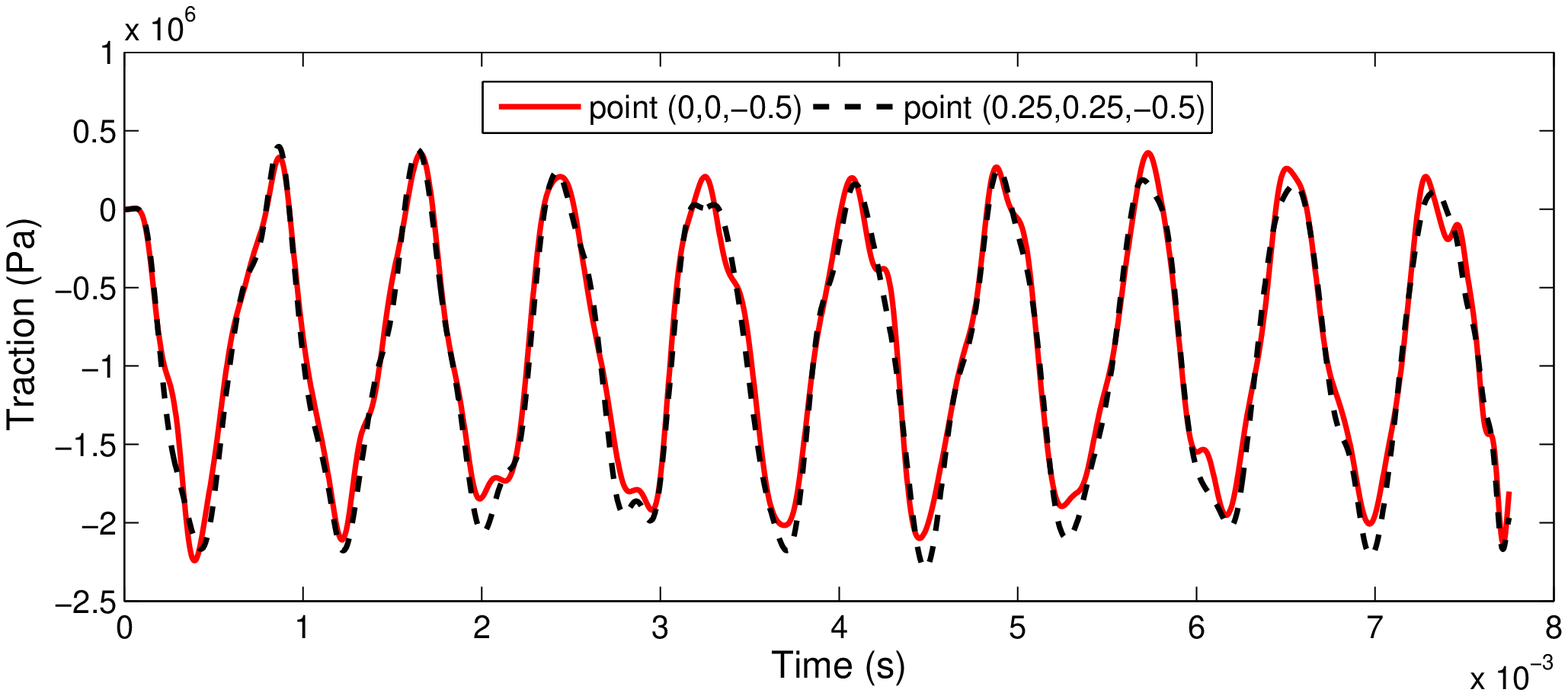,width=0.8\textwidth}\\
(b) Traction
\caption{The displacement and traction responses for a few selected points} \label{fig-NE3-disp_tract}
\end{figure*}

\section{Summary}\label{S-Conclution}

Fast solvers are indispensable for large-scale wave propagation analysis. Recently, a fast frequency domain pFFT BEM incorporated with the exponential window technique was developed for solving transient elastodynamic problems \cite{XYCZ12}.
It has been shown that this method is stable and general in the sense that: (1) the method can be used to obtain accurate time responses for an arbitrarily chosen time period; (2) the method is applicable to solving problems with damping, either small or large, as long as the damping term is linear and the resulting frequency-domain equation has an equivalent integral formulation

In this work, the computational efficiency of the method in \cite{XYCZ12} is further improved via three approaches:
\begin{enumerate}
  \item \emph{Introducing large damping} via the exponent window function. It can be proved that the damping actually plays the role of \emph{preconditioner}, denoted as the artificial damping preconditioner (ADP) in this paper. Applying a large damping can in general improve the conditioning of the BEM matrices, thus reducing the total number of iterations required to achieve convergence. Numerical results show that the reduction is around 25\% for $\kappa=4$ compared with $\kappa=2$.

  \item \emph{Using frequency windowing technique} in Fourier inversion. Exponential function with a large damping coefficient would cause unacceptable oscillations in the later time responses.  The oscillations stem from the Gibbs phenomenon which is inherent in expanding non-periodic functions using Fourier series. The frequency domain windowing technique is an effective way to abate the Gibbs phenomenon and is employed in the Fourier inversion of the frequency domain BEM results. It is found that the well-known Blackman and Hanning windows perform equally well in removing the Gibbs oscillations in the BEM solutions. The performance of the Hanning window is demonstrated by numerical results in Section \ref{S-NE}.
      The time-domain damping in conjunction with the frequency-domain windowing techniques results in a modified Fourier transform method for elastodynamic wave propagation simulation.

  \item \emph{Using solution extrapolation method (SEM)} for obtaining good initial guess. In this method, the initial guess for the current linear system is obtained from the solutions of the linear systems corresponding to several previous sampling frequencies via extrapolation. The aim is to reduce the number of iterations for iterative solvers. Numerical results indicate that the reduction can be up to 15\% and is generally problem dependent.
\end{enumerate}

The effectiveness of the above strategies is demonstrated by three examples. The largest model has nearly 0.7 million DOFs, and is simulated on a workstation with a Xeon 3.0GHz CPU and 30GB RAM.

\begin{acknowledgements}
JX and LW were supported by the NSFC under Grant 11102154 and 11074201, and the New Teacher Fund for Doctor Station from the Chinese Ministry of Education under Grant 20106102120009.
WY was supported by Hong Kong Research Grants Council under Competitive Earmarked Research Grant 621411.
\end{acknowledgements}



\end{document}